# Anisotropy in the electronic screening of oxygen lattice modes in YBa$_2$Cu$_3$O$_{6.95}$

R. J. McQueeney [1*], T. Egami [2], J.-H. Chung [2], Y. Petrov [3§], M. Yethiraj [4], M. Arai [5], Y. Inamura [5], Y. Endoh [6], C. Frost [7] and F. Dogan [8]

[1] Los Alamos National Laboratory, Los Alamos, New Mexico 87545 USA
[2] Department of Materials Science and Engineering, University of Pennsylvania, Philadelphia, PA 19104 USA
[3] Department of Physics and Astronomy, University of Pennsylvania, Philadelphia, PA 19104 USA
[4] Oak Ridge National Laboratory, Oak Ridge, TN 37831, USA
[5] Institute of Materials Structure Science, KEK, Tsukuba 305-0801, Japan
[6] Institute for Materials Research, Tohoku University, Sendai 980, Japan
[7] Rutherford Appleton Laboratory, Didcot, Oxon, OX11 0QX, UK
[8] Department of Materials Science, University of Washington, Seattle, WA 98195, USA

Inelastic neutron scattering data from a twinned single-crystal of YBa$_2$Cu$_3$O$_{6.95}$ are presented that show a distinct $a$-$b$ plane anisotropy in the oxygen vibrations. The Cu-O bond-stretching type phonons are simultaneously observed along the $a$ and $b$ directions due to a 4 meV splitting arising from the orthorhombicity. The present results show the bond-stretching branch along $b$ (parallel to the chain) has a continuous dispersion, while the branch along $a$ is discontinuous, suggesting a possibility of short-range cell-doubling along $a$. Furthermore, the LO mode along $a$ is split in energy from its TO partner at non-zero $q$-vectors, while the $b$ mode is not. These results imply strong anisotropy in the electronic screening and a one-dimensional character in underlying charge fluctuations.

In recent years, evidence has been building that the lattice may play a non-trivial role in the transport and superconducting properties of high-temperature superconductors. In particular, certain modes (the high-energy Cu-O bond-stretching modes) show strong softening with doping, and their frequencies are strongly and abruptly reduced when crossing from the insulating to the metallic phase as a function of doping [1-4]. Photoemission also suggests strong electron-phonon interaction for these modes [5,6]. The compendium of measurements seem to support an anisotropic, locally inhomogeneous charge model, which may be related to the stripe model [7,8]. In the present Letter, we show that the dispersion of Cu-O bond-stretching phonons in orthorhombic YBa$_2$Cu$_3$O$_{6.95}$ (YBCO) are very different along the $a$ and $b$ directions, with the dispersion along $a$ being discontinuous half way to the zone boundary. This is also the modulation wavevector direction for incommensurate spin fluctuations observed in untwinned single-crystals [9] establishing a connection between the Cu-O bond-stretching modes and the stripe structure.

Inelastic neutron scattering measurements were performed on a twinned single-crystal of YBa$_2$Cu$_3$O$_{6.95}$ weighing ~100 grams. Measurements were focused on the Cu-O bond-stretching type phonons along the $a/b$-direction in the twinned crystal. Untwinned single-crystals of sufficient size for phonon measurements are not available. The $a$-direction is called $\Sigma$ and the $b$-direction $\Delta$ according to standard notation [10]. YBCO has a distinct orthorhombicity due to the presence of CuO chains running along the $b$-direction. In addition, the YBCO crystal structure contains two CuO$_2$ planes separated by 3.38 Å. The high-energy bond-stretching phonons are oxygen modes polarized in the CuO$_2$ plane and can have either in-phase displacements between the two planes ($\tau_1$ symmetry) or out-of-phase displacements ($\tau_3$ symmetry). Previous measurements have shown that these phonon modes have strongly reduced frequencies compared to the insulating parent compound YBa$_2$Cu$_3$O$_6$ between $q = (\pi/2a, 0, 0)$ and $(\pi/a, 0, 0)$ and the corresponding wavevectors along $b$ [4].

---

[*] email: mcqueeney@lanl.gov
[§] Present address: Gatsby Neuroscience Computational Unit, University College London, London WC1N 3AR, UK

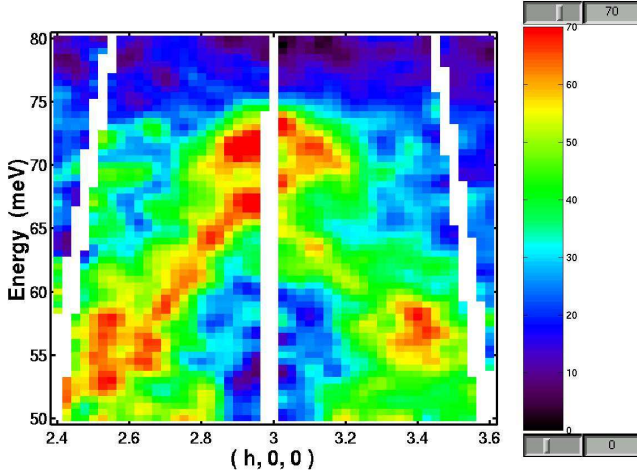

FIG. 1. Neutron scattering intensity map by pulsed neutron inelastic scattering performed at the ISIS using the MAPS spectrometer at T=110 K. Data are shown as a function of energy transfer and momentum transfer ($q$), in the range $2.5 < q_x < 3.5$, $-0.05 < q_y < 0.05$, in the unit of reciprocal vector, $2\pi/a$. The data were divided by $q^2$, after subtracting the background, since the phonon intensity is proportional to $q^2$. The data were taken at T=110 K. White lines are due to gaps in the detector coverage.

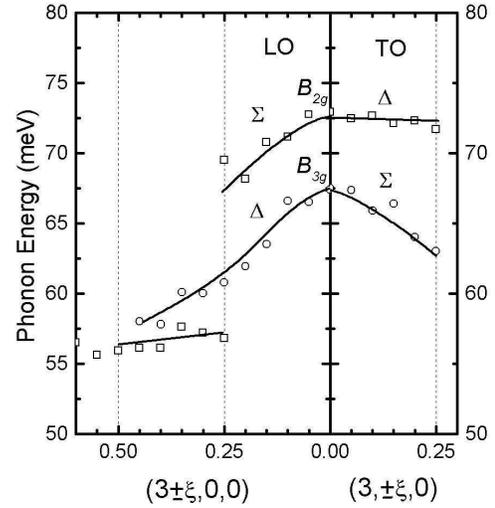

FIG. 2. The dispersion of the Cu-O bond-stretching phonon modes in YBa$_2$Cu$_3$O$_{6.95}$ at T=7 K, determined by curve fitting to the MAPS data. The shift between the modes seen in Fig. 1 is corrected here. The left-hand side describes the LO phonons and the right-hand side the TO phonons. The branch that is denoted $B_{3g}$ at the zone-center (o) has polarization along the $b$ axis (parallel to the chain). This mode is smoothly dispersing and TO ($\Sigma$) and LO ($\Delta$) are nearly degenerate. On the other hand, the $B_{2g}$-$\Sigma$ branch that has polarization along $a$ (□) is discontinuous at the middle of the zone, and joins the $B_{3g}$-$\Delta$ mode at the zone-edge. The LO frequency of the $B_{2g}$-$\Sigma$ branch is significantly lower than the TO ($\Delta$) frequency for non-zero $q$. Lines are guide to the eye.

Measurements were performed by neutron time-of-flight spectroscopy on the MAPS spectrometer at the ISIS facility of Rutherford Appleton Laboratory using an incident energy of 118 meV. The sample orientation had the $a/b$-axis perpendicular to the scattering plane and the $c$-axis was rotated 41° away from the incident beam direction. MAPS data were analyzed using the Mslice program. Additional measurements on the same crystal were performed by triple-axis spectrocopy on the HB-2 and HB-3 instruments at the High-Flux Isotope Reactor at Oak Ridge National Laboratory. For the triple-axis measurements the final energy was fixed at 14.87 meV and the sample-to-detector horizontal collimation elements were 48'-60'-40'-120'.

Due to the presence of the Cu-O chains, some degree of anisotropy is expected in the lattice dynamics in the CuO$_2$ plane from orthorhombicity. At the Brillouin zone center, Raman scattering on untwinned single-crystals has identified that the $B_{2g}$ ($a$-axis polarization) and $B_{3g}$ ($b$-axis) modes which connect to the ($\tau_3$) bond-stretching modes are split by 4 meV [11]. We note that the in-plane displacements of the chain oxygen modes are not Raman active, except perhaps weakly due to disorder, therefore the above mode assignments must be correct. This splitting is also confirmed by neutron scattering on twinned crystals [12] and the results presented here. Neutron scattering shows the same splitting for the $B_{2u}$ and $B_{3u}$ modes, which connect to the $\tau_1$ modes [13]. The $\tau_3$ modes around the reciprocal lattice point (3,0,0) (in the unit of reciprocal lattice vectors) are shown in Fig. 1 where the $B_{2g}$ and $B_{3g}$ modes are observed at 72 and 68 meV, respectively [14]. Note that the dispersion maximum for the $B_{2g}$ mode occurs around (3.02,0,0) with the provisional lattice constant of 3.85 Å, while that for the $B_{3g}$ mode is located at (2.96,0,0). This is because $a$ is smaller than $b$ by 2%, confirming that each mode propagates along a different orthorhombic axis. The phonon splitting due to orthorhombicity is consistent with the higher frequency mode having a shorter unit cell distance.

The splitting of the dispersions of the $\Delta$ and $\Sigma$ modes allow to follow each mode independently, in spite of the fact that the crystal is twinned. As shown in Fig. 2 the lower energy $\Delta$ branch disperses strongly from 68 meV down to 55 meV near the zone boundary Y-point (0, $\pi/b$,0) where it mixes with an oxygen bond-bending branch of the same symmetry. The $\Sigma$ branch, on the other hand, disperses from 72 meV to 68 meV at the ($\pi/2a$,0,0) point where the intensity rapidly disappears. The intensity of the $\Sigma$ branch reappears near the zone boundary at ~55 meV. This discontinuous nature of the bond-stretching phonon branch can arise from strong electron-lattice coupling and may be understood as a dynamic cell-doubling [1,15,16]. Here we show that this gapped behavior of the bond-stretching branch occurs only for bond-stretching phonons propagating along the $a$-direction, while the $b$-direction has a continuous dispersion. This suggests that the formation of short-ranged charge fluctuations with length scale $2a$ occur only along the $a$-direction.

The one-dimensionality of the electron-lattice interaction is further enforced by examining the transverse modes (with $\tau_2$ symmetry) emanating from the $B_{2g}$ and $B_{3g}$ phonons. The behavior of the transverse modes is shown in the right-hand side of Fig. 2. The transverse mode along $\Delta$ (with polarization along the $a$-direction) remains dispersionless at 72 meV until at least $(0, \pi/2b, 0)$, the extent of our measurement range. The transverse mode along $\Sigma$ has a dispersion similar to the longitudinal bond-stretching $\Delta$ mode. As expected for a metal, the LO/TO splitting is zero at the zone center due to electronic screening. The LO/TO splitting remains very small for oxygen displacements along the $b$-direction, however, it increases with $q$ along $a$. This implies that the phonons polarized in the $b$-direction remain well screened and more metallic than those polarized in the $a$-direction. In other words, the electronic dielectric function, $\varepsilon_{el}(q, \omega)$, is different along $a$ and $b$ and the low-frequency charge fluctuations in YBCO are very anisotropic. Moreover, the LO frequency is *lower* than the TO frequency for non-zero-$q$, which is anomalous in the spirit of the Lyddane-Sachs-Teller relationship [17]. The effect of electronic screening due to charge fluctuations was investigated via the calculation of $\varepsilon_{el}(q, \omega)$ by Tachiki and Takahashi [18]. They have shown that overscreening due to the vibronic state, in which charge and lattice vibrate at the same frequency with some phase shift, results in the similar softening of the LO mode for $q \neq 0$.

The dispersions of the LO bond-stretching phonon branch were found to show no apparent temperature dependence, but the phonon intensity was found to show significant change with temperature, beyond what is expected for thermal excitation. Figure 3 shows the difference in intensity between 7 K and 110 K. There is a transfer of the phonon intensity from a position halfway to the zone boundary (in the energy range from 55 - 70 meV) to the zone-center (70 - 75 meV) and zone-boundary (50 - 55 meV). A very similar difference pattern was observed also in the triple-axis data [19]. This transfer in spectral weight can be expressed in terms of the intensity at $q=(\pi/2a,0,0)$ as the difference in the average intensity between the Range 1 (51 - 55 meV) and the Range 2 (56 - 68 meV). Fig. 4 shows the difference, $I(1) - I(2)$, obtained with the triple-axis spectrometer. It grows below $T_c$, just as the superconducting order parameter.

The anisotropy of the phonon dispersion and implied anisotropy of electronic screening suggests that charges have low-frequency fluctuations which are strongly anisotropic. They could possibly originate from an array of oriented one-dimensional metallic channels separated by less metallic regions, as in the stripe model [8,20,21]. Then, ionic displacements along the metallic channels are well screened, whereas displacements perpendicular to the chan-nels are poorly screened due to an energy gap for holes jumping out of the channels. The incommensurate spin fluctuations signify the presence of such charge and spin modulation [9]. Inelastic neutron scattering measurements with untwinned crystals of YBCO show that the spin modulation occurs primarily along the $a$-direction, inferring that stripes are oriented along $b$ [22]. This observation is entirely consistent with metallic screening channels being oriented along $b$. What remains unclear is the spacing of the charge channels, which is more likely to be $2a$ from analysis of the phonon data [1], but assumed to be $4a$ by extrapolation of the diffraction measurements of pinned stripe materials to dynamic stripes [8,20].

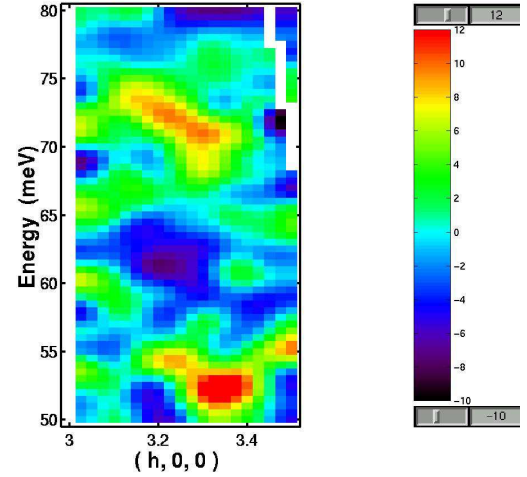

FIG. 3. Difference in the scattering intensity between 7 K and 110 K, along $h$, determined by pulsed neutron scattering with the MAPS at the ISIS. To increase the statistics the data for q=2.5-3.0 and q=3.0-3.5 were averaged after mirror operation. Below $T_c$ the intensity increases in the energy ranges 50 - 55 meV and 67 - 75 meV, while decreases in the energy range 55 - 67 meV.

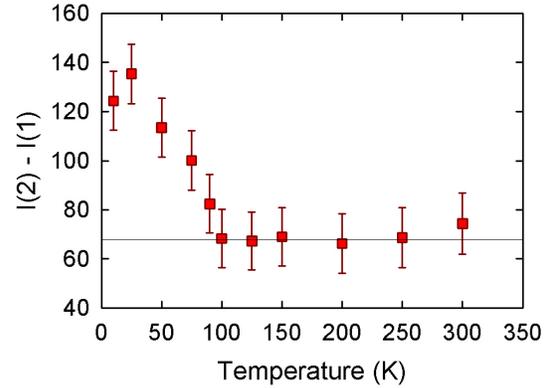

FIG. 4. Temperature dependence of the difference in the neutron scattering intensity at $q=(3.25,0,0)$ between the energy range 1 (51-55 meV) and range 2 (56-68 meV). The plot shows $I(1)-I(2)$, determined by the triple-axis spectrometer. The superconducting transition temperature of this sample is 93 K; the data show a striking resemblance to the temperature dependence of the superconducting order parameter.

The temperature dependence shown in Fig. 4 suggests that the gapped behavior of the Cu-O bond-stretching mode is directly related to superconductivity. The changes in the intensity with temperature shown in Fig. 3 indicate that below $T_c$ even the $\Delta$ branch along the $b$ direction develops a weak gap around 62 meV, so that the dispersion gap becomes two-dimensional. This may be related to the two-dimensional nature of superconducting correlation [21]. It is important to note that the present result is consistent with the recent ARPES results on an untwined YBCO crystal [24] which show a strong anisotropy in the electronic structure and a higher density of superconductive condensate in the direction perpendicular to the chain. These results suggest the possibility of superconductivity via the overscreening mechanism as suggested by Tachiki and Takahashi [18], the phonon umklapp mechanism by Castro Neto [23], or the coupled spin-ladder mechanism of Sachdev [16]. For La$_{1.85}$Sr$_{0.15}$CuO$_4$ (LSCO) both the dispersion with a jump [1] as well as a continuous dispersion [25] were observed, suggesting that both of them exist in LSCO just as in YBCO [26]. Then the anisotropy in the screening may be a common feature of the cuprate superconductors, not a byproduct of the chain. In YBCO the orthorhombicity and the presence of the chain must favor the alignment of stripes in one direction throughout a twin domain, while in LSCO two types of domains with stripes along $x$ and $y$ must coexist. For large YBCO samples where twinning tends to obscure this anisotropy, the energy splitting of the bond-stretching phonons is fortuitous, and allows one to effectively observe a single domain lattice response for these strongly coupled modes. From the observations of the anisotropy of the lattice dynamics in this Letter and spin dynamics and charge dynamics in the literature, it appears that all aspects of the low frequency dynamics are strongly anisotropic, as suggested by the stripe model. This places a stringent constraint to theories of high-temperature superconductivity.

The authors would like to thank H. A. Mook, M. Tachiki, A. R. Bishop, A. Bussmann-Holder, A. H. Castro Neto, S. Sachdev, S. Kivelson, and P. Allen for helpful discussions. This work was supported by the U.S. Department of Energy under contract number W-7405-Eng-36 with the University of California and the National Science Foundation under grant DMR96-28134 to the University of Pennsylvania.